\documentclass[aps,prc,floatfix,nofootinbib,preprintnumbers,superscriptaddress,longbibliography,notitlepage]{revtex4-1}

\usepackage{epsfig}
\usepackage{graphicx}
\usepackage{stmaryrd}
\usepackage{amssymb,tabularx,dcolumn}
\usepackage{supertabular,ltxtable}
\usepackage{longtable}
\usepackage{amsmath}
\usepackage{amstext}
\usepackage{float}
\usepackage{color}
\usepackage{bm}
\usepackage{hyperref}
\usepackage{lipsum}
\hypersetup{breaklinks=true,colorlinks=true,linkcolor=blue,citecolor=blue,filecolor=magenta,urlcolor=cyan}
\usepackage{mathtools}
\usepackage{multirow}
\usepackage{makecell}
\usepackage[utf8]{inputenc}
\usepackage{tabu}
\usepackage{braket}
\usepackage{subfigure}
\usepackage{enumitem}

\usepackage[dvipsnames]{xcolor}
\definecolor{pastelgray}{rgb}{0.81, 0.81, 0.77}
\definecolor{beaublue}{rgb}{0.9, 0.9, 0.93}

\newcommand{\PRLsep}{\noindent\makebox[\linewidth]{\resizebox{0.5\linewidth}{1pt}{$\bullet$}}\bigskip}

\newcommand{\NME}{$M_{0 \nu}$}

\makeatletter
\def\@bibdataout@aps{%
\immediate\write\@bibdataout{%
@CONTROL{%
apsrev41Control%
\longbibliography@sw{%
    ,author="08",editor="1",pages="1",title="0",year="1"%
    }{%
    ,author="08",editor="1",pages="1",title="",year="1"%
    }%
  }%
}%
\if@filesw \immediate \write \@auxout {\string \citation {apsrev41Control}}\fi
}
\makeatother

\begin{document}

\title{Towards Precise and Accurate Calculations of Neutrinoless Double-Beta Decay: Project Scoping Workshop Report}

\author{V.~Cirigliano}
\affiliation{Institute for Nuclear Theory, University of Washington, Seattle, WA 98195, USA}

\author{Z.~Davoudi}
\affiliation{
 	Department of Physics and Maryland Center for Fundamental Physics, 
University of Maryland, College Park, MD 20742, USA.}

\author{J.~Engel}
\email{engelj@physics.unc.edu}
\affiliation{Department of Physics, University of North Carolina, Chapel Hill, NC 27514, USA}

\author{R.J.~Furnstahl}
\affiliation{Department of Physics, Ohio State University, Colubmus, OH 43210, USA}

\author{G.~Hagen} 
\affiliation{Physics Division, Oak Ridge National Laboratory,
Oak Ridge, Tennessee 37831, USA}
\affiliation{Department of Physics and Astronomy, University of Tennessee,
Knoxville, Tennessee 37996, USA}

\author{U.~Heinz}
\affiliation{Department of Physics, Ohio State University, Colubmus, OH 43210, USA}

\author{H.~Hergert}
\affiliation{FRIB Laboratory and Department of Physics and Astronomy, Michigan State University, East Lansing, Michigan 48824, USA}

\author{M.~Horoi}
\affiliation{Department of Physics, Central Michigan University, Mount Pleasant, MI 48859, USA}

\author{C.W.~Johnson}
\affiliation{Department of Physics, San Diego State University, San Diego, CA 92182, USA}

\author{A.~Lovato}
\affiliation{Physics Division, Argonne National Laboratory, Lemont, Illinois 60439, USA}
\affiliation{Computational Physics Division, Argonne National Laboratory, Argonne, IL 60439, USA}
\affiliation{INFN-TIFPA Trento Institute of Fundamental Physics and Applications, 38123 Trento, Italy}

\author{E.~Mereghetti}
\affiliation{Theory Division, Los Alamos National Laboratory, Los Alamos, NM 87544, USA}

\author{W.~Nazarewicz}
 \email{witek@frib.msu.edu}
\affiliation{FRIB Laboratory and Department of Physics and Astronomy, Michigan State University, East Lansing, Michigan 48824, USA}

\author{A.~Nicholson}
\affiliation{Department of Physics, University of North Carolina, Chapel Hill, NC 27514, USA}
\affiliation{Nuclear Science Division, Lawrence Berkeley National Laboratory, Berkeley, CA 94720, USA}

\author{T.~Papenbrock}
\affiliation{Department of Physics and Astronomy, University of
  Tennessee, Knoxville, Tennessee 37996, USA}
\affiliation{Physics Division, Oak Ridge National Laboratory, Oak
  Ridge, Tennessee 37831, USA}

\author{S.~Pastore}
\affiliation{Department of Physics and McDonnell Center for the Space Sciences, Washington University in Saint Louis, Saint Louis, MO 63130}

\author{M.~Plumlee}
\affiliation{Industrial Engineering and Management Sciences, Northwestern University, Evanston, Illinois 60208, USA}
\affiliation{NAISE, Northwestern University, Evanston, Illinois 60208, USA}

\author{D.R.~Phillips}
\email{phillid1@ohio.edu}
\affiliation{Department of Physics and Astronomy and Institute of Nuclear and Particle Physics, Ohio University, Athens, OH 45701, USA}

\author{P.E.~Shanahan}
\affiliation{Center for Theoretical Physics, Massachusetts Institute of Technology, Cambridge MA 02139, USA}

\author{S.R.~Stroberg}
\affiliation{Physics Division, Argonne National Laboratory, Lemont, Illinois 60439, USA}

\author{F.~Viens}
\affiliation{Department of Statistics and Probability, Michigan State University, East Lansing, MI 48824, USA}

\author{A.~Walker-Loud}
\affiliation{Nuclear Science Division, Lawrence Berkeley National Laboratory, Berkeley, CA 94720, USA}

\author{K.A.~Wendt}
\affiliation{Nuclear and Chemical Sciences Division, Lawrence Livermore National Laboratory, Livermore, CA 94550}

\author{S.M.~Wild}
\affiliation{Mathematics and Computer Science Division, Argonne National Laboratory, Lemont, Illinois 60439, USA}
\affiliation{NAISE, Northwestern University, Evanston, Illinois 60208, USA}

\date{\today}

\maketitle

\clearpage\newpage
\tableofcontents

\clearpage\newpage
\section*{Executive Summary}

\noindent
This white paper is an outgrowth of a National Science Foundation (NSF) Project Scoping Workshop, the purpose of which was to assess the current status of calculations for the nuclear matrix elements governing neutrinoless double-beta decay and determine if more work on them is required. The recent effort to define the United States' role in ton-scale experiments, together with the conclusion in 2021 of the Department of Energy (DOE) Topical Collaboration on neutrinoless double-beta decay, made such an exercise extremely timely.
 
The main conclusions of the workshop can be summarized as follows:
\begin{itemize}
    \item Neutrinoless double-beta decay of nuclei is an important window into the physics of neutrinos. It could be the first lepton-number violating process ever observed. As such, it would provide key insights into physics beyond the Standard Model---in particular how the matter-anti-matter asymmetry of the universe arose.
    
    \item Much progress on the theory of neutrinoless double-beta decay of nuclei has been made over the last five years. An end-to-end set of effective field theories (EFTs) that shows how to evolve the physics of the decay from the scale at which lepton number is violated (possibly much larger than TeV)
    down to the scales relevant for nuclei has been developed. Lattice quantum chromodynamics (LQCD) calculations of the process in the two-nucleon (NN) system are being set up, and the first ab initio many-body calculations of neutrinoless double-beta decay matrix elements {\NME} have been carried out.
    
    \item Much remains to be done before theory can successfully complement the large experimental effort to observe neutrinoless double-beta decay and measure its rate.
    Both the accuracy and precision of LQCD and ab initio nuclear many-body calculations need to be improved if crisp statements about experimental observation or non-observation are to be made. The uncertainty in nuclear many-body calculations remains largely unquantified, making it difficult to interpret the significant differences predicted by different approaches for the rate of neutrinoless double-beta decay in candidate nuclei.
    
    \item Uncertainty quantification is thus crucial to future progress. Better assessment of both the parametric uncertainty and the model uncertainty in predictions of neutrinoless double-beta decay matrix elements is needed. The tools for the former exist, and methodology for the latter is under development. The ab initio calculation of a variety of nuclear observables 
    related to neutrinoless double-beta decay can help establish and reduce the uncertainty in {\NME} that arises from the complexity of the nuclear many-body problem. 
\end{itemize}

\section{Introduction}
\noindent
In recent years the search for new fundamental physics, for the forces and
particles that underlie the Standard Model, for the explanation of 
the excess of matter over antimatter and similar mysteries, and for
the sources both of symmetries and their violation, has moved
increasingly to low-energy experiments.  Among the most visible and promising
are those that seek to observe neutrinoless double-beta ($0\nu\beta\beta$)
decay, a process in which two neutrons inside an atomic nucleus turn into
protons, emitting two electrons and no neutrinos.  An observation of this
process would show that lepton number (L) is not conserved and that the neutrino
mass has a Majorana component, implying that the mass eigenstates are
self-conjugate~\cite{Schechter:1981bd}.  Observation of $0\nu\beta\beta$ decay
would thus provide crucial information about neutrino mass
generation~\cite{Minkowski:1977sc,Mohapatra:1979ia,GellMann:1980vs}, and suggest
that the matter-antimatter asymmetry in the universe originated in
leptogenesis~\cite{Davidson:2008bu}. The major implications of an observation
made the construction of a ton-scale $0\nu\beta\beta$-decay experiment the top
priority for new projects in the 2015 NSAC Long Range
Plan~\cite{Geesaman:2015fha}, which set the decadal priorities for nuclear
physics. The anticipated investment is in the range of 250-400 million dollars.  

Smaller experiments already put stringent limits on the decay rate
\cite{Gando:2012zm,Agostini:2013mzu,Albert:2014awa,Andringa:2015tza,KamLAND-Zen:2016pfg,Elliott:2016ble,Agostini:2017iyd,Aalseth:2017btx,
Albert:2017owj,Adams2022,Agostini:2018tnm, Azzolini:2018dyb}, e.g.
$T^{0\nu}_{1/2}>2.3\times 10^{26}$yr for the decay of ${}^{136}$Xe
\cite{KamLAND-Zen:2022tow}. The next very few years will see stricter limits from
experiments---such as LEGEND-200, CUORE, KamLAND-Zen 800, and SNO+---that are currently 
operating or under construction.
On a slightly longer time scale, ton-scale experiments
\cite{armstrong2019cupid,abgrall2021legend,adhikari2021nexo,adams2020sensitivity,KamLAND-Zen:2022tow,SNO:2021xpa}
based on $^{76}$Ge, $^{100}$Mo, $^{136}$Xe, and perhaps other isotopes will come on line.  The
goal of these large experiments is the ability to detect any decay caused by the
exchange of light Majorana neutrinos if the neutrino mass hierarchy is inverted
(i.e., if the neutrino with the largest electron-flavor component is the
heaviest), as well as increased sensitivity to decay caused primarily by the
exchange of other still-hypothetical particles.  

In order to extract the effective light-neutrino Majorana mass $m_{\beta \beta}
\equiv | \sum_{i} U_{ei}^2 m_i|$ (with $m_i$ the mass of the neutrino mass eigenstate $i$ and $U_{ei}$ the elements of the Pontecorvo-Maki-Nakagawa-Sato matrix) from any of these impressive experiments, one
needs nuclear matrix elements (denoted by $M_{0\nu}$) of the decay operators. 
The degree to which ton-scale experiments will be sensitive to decay
caused by the exchange of inverted-hierarchy light neutrinos
depends on these nuclear matrix elements, as does the the extent to which 
experiments in more than one isotope will prove useful.   The nuclear matrix 
elements suffer at present from sizable
uncertainties~\cite{Engel:2016xgb}.  \textit{Their accurate computation, with a
quantified uncertainty, is therefore a task of the greatest importance.}

The need for precise nuclear matrix elements is in fact more general than the
notion that the exchange of light Majorana neutrinos causes $0 \nu \beta \beta$
decay.  That idea is based on the assumption that lepton-number violation (LNV) originates at very high
energies and manifests itself in the decay through the ``high-scale seesaw,''
which leaves Majorana neutrino masses as its only remnant at low
energies.  If that is indeed the case, $0\nu\beta\beta$ decay and
neutrino-oscillation experiments will together tell us most of what we can
learn.  High-scale LNV is only one scenario, however, and even in the
restricted class of seesaw models, it applies only if the new particles are very
heavy right-handed neutrinos.  In many Beyond-the-Standard-Model (BSM) scenarios, other
lower-scale sources of LNV can also induce $0\nu\beta\beta$ decay.  In left-right symmetric
models, for example, heavy neutrinos and charged scalars with 
TeV-scale 
masses can be exchanged. 
In other scenarios there may be light right-handed (sterile) neutrinos with
masses much lower than the electroweak scale.  The large number of ways in which
lepton number could be violated (see, e.g., Ref.~\cite{Rodejohann:2011mu} for a review)
means that ton-scale searches for $0\nu\beta\beta$ decay have a significant
discovery potential beyond the inverted-hierarchy high-scale seesaw.   Each kind
of LNV leads to its own set of transition operators, the nuclear matrix elements
of which must be calculated.  If the calculations are sufficiently accurate,
we can assess the sensitivity of the generation of experiments now coming online  to various
kinds of LNV. We can also provide a subsequent generation of experiments with information on how best to
narrow the range of
possibilities for LNV and neutrino-mass generation through measurements of
single-electron spectra, electron angular distributions, and the isotopic
dependence of the decay rate. 

\begin{figure}
    \centering
\includegraphics[scale=0.675]{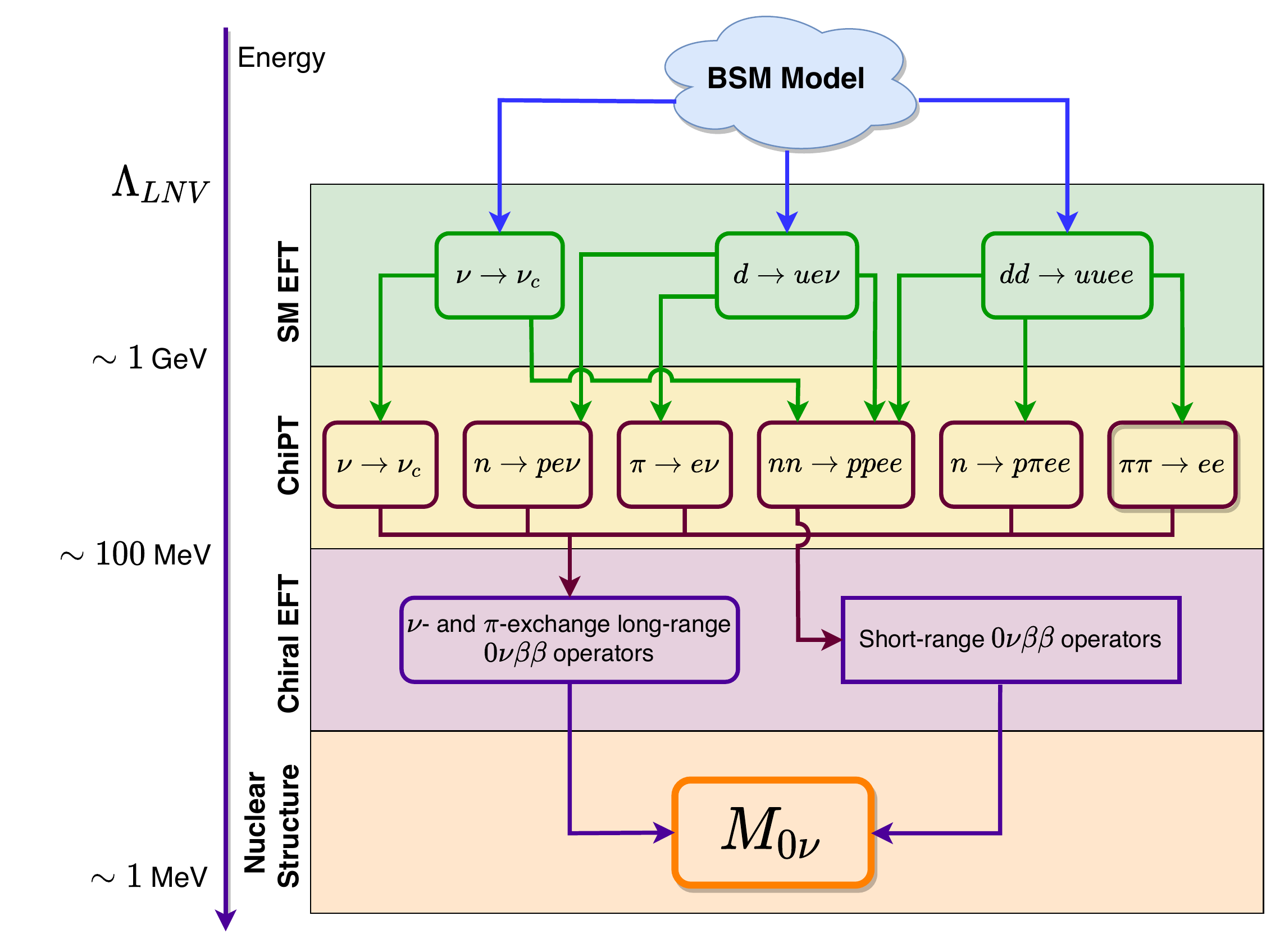}
    \caption{A ``tower of theories'' leading to the computation of the nuclear matrix elements $M_{0\nu}$ that control the rate of $0\nu\beta\beta$ decay.  At the highest level, above the cutoff $\Lambda_{LNV}$ for all effective theories, is the ultimate BSM LNV.  It manifests itself at the quark and gluon level through Standard-Model EFT, at the nucleon and pion level through chiral perturbation theory (ChiPT), at the nucleon-only level (i.e., with pions no longer part of the Hilbert space, but instead accounted for in multi-nucleon operators) through chiral EFT, and at the nuclear level through the techniques of nuclear-structure theory.
    Figure adapted from Ref.~\cite{Cirigliano:2018yza}. 
    }
    \label{fig:tower}
\end{figure}

The ability to compute all the relevant nuclear matrix elements requires work at
widely separated energy scales, from the high energies at which LNV originates
all the way down to nuclear energies, and the ability to bridge the scales.
EFT provides the bridge by expanding observables and
Lagrangians in the ratios of the important energy scales. In reality the calculation is done via a series of EFTs---a connected set of bridges rather than a single one; see Fig.~\ref{fig:tower} for an illustration. The SM EFT allows us to encode the effects of different LNV mechanisms in operators involving neutrinos, electrons, and $d$ and $u$ quarks, thereby taking us from the TeV scale to the scale of quark confinement at around 1 GeV. Converting these operators into hadronic operators that are organized through chiral perturbation theory requires non-perturbative input from LQCD. Chiral-perturbation-theory operators are then used to derive operators in a nucleons-only Hilbert space; following that step, the operators can be used in many-body calculations of nuclei. In combination, the bridges deliver us a separate set of chiral EFT $nn \to pp$ transition operators for each LNV
source that are to be used in nuclear many-body calculations.  The combination
of SM EFT, LQCD, chiral EFT, and ab initio (first-principles) nuclear many-body methods,
each of which has the ability to control uncertainty, therefore provides a path---the only path, in fact---toward the reliable estimation of
uncertainties in {\NME}.

Chiral EFT is key to the progress made to this point, and to future efforts to quantify uncertainties. Chiral EFT~\cite{Weinberg1990,Weinberg1991,Epelbaum2008,Hammer2019} is the extension of chiral perturbation theory to the few-nucleon problem.  Just as with chiral perturbation theory, chiral EFT is organized as an expansion in powers of $p/\Lambda$ or $m_\pi/\Lambda$, where $p$ is a typical nucleon momentum, $m_\pi$ is the pion mass and $\Lambda$ is the theory's ``breakdown scale'' of about 500 MeV.  But chiral EFT is not a perturbative theory, because it has to account for nuclear binding. Although discussions of exactly how to do that continue (see, e.g., \cite{vanKolck2020}) chiral EFT has the virtue of delivering consistent 
 nuclear forces and $0 \nu \beta \beta$ operators up to a given order in the chiral EFT expansion. Even better, these operators include the consequences of QCD's chiral symmetry, e.g., connections between pionic operators and the axial current that governs beta decay. Perhaps most significantly for the purposes of this document, chiral EFT permits estimation of the uncertainty associated with the model of the nuclear force and the interactions that govern $0 \nu \beta \beta$ decay. A $k$th order chiral EFT calculation should have a fractional error of $O(\{p,m_\pi\}^{k+1}/\Lambda^{k+1})$. It follows that different implementations of chiral EFT---different orders of the calculation, different regulator choices---should give answers that are consistent with one another once this error estimate is taken into account. Bayesian techniques have recently been employed to quantify this error~\cite{Furnstahl2015}, and show that---provided the chiral EFT calculation is implemented carefully---the error estimate provides a good account of the predictive accuracy of chiral EFT in light nuclei~\cite{LENPIC2020,LENPIC2022}. Chiral-EFT forces and operators therefore provide the starting point for ab initio calculations that use the nuclear many-body methods described below.

The nuclear-theory community has   made significant progress, at all the levels in this tower of EFTs, toward more accurate calculations of {\NME}. But this progress has in part served to confirm that there are $O(1)$ uncertainties in {\NME}. These uncertainties (unless reduced) will prevent us from learning about the sources of LNV, even if
several experiments detect the process.

There is therefore still much to do. In particular:
\begin{itemize}
\item The $0\nu\beta\beta$ transition operators used in
nuclear-structure physics are now written in terms of ``low-energy constants''
(LECs) that multiply terms in the chiral-EFT Lagrangian that is used at the hadronic scale.  In chiral EFT, the LECs multiplying the terms at the lowest orders are thus the most important.  Previously unrecognized
LECs associated with zero-range $nn \to pp$ transition operators appear even at
leading order in the $0\nu\beta\beta$ piece of the chiral Lagrangian, for both light-Majorana neutrino
exchange~\cite{Cirigliano:2018hja}
and TeV scale LNV~\cite{Cirigliano:2018yza}.  We must improve our
knowledge of these LECs, both by relating $0\nu\beta\beta$ decay to other $\Delta I
=2$ processes and by direct calculation within LQCD.  

\item To use the results of EFT and LQCD in the computation of nuclear matrix
elements---that is, to use the chiral-EFT Hamiltonians and transition operators that these methods supply in many-body calculations---we need to improve
ab initio methods.  The improvement will involve an increase in accuracy, 
the use of a wide range of chiral-EFT Hamiltonians (to allow
uncertainty quantification), and a careful analysis of the way such methods
employ the EFT operators.  The first two of these will require, in addition to
analytic work, more efficient use of our best supercomputing resources.
Existing codes and their extensions will need to be reworked to leverage
accelerators such as GPUs.  Benchmarking with methods that are known to give very
accurate results (so-called ``quasi-exact'' methods that have thus far been restricted by complexity to light nuclei) is also important.

\item At both the hadronic and nuclear scales, we need a consistent and unified
quantification of uncertainties.  We must be able to both 
propagate parameter uncertainties to observables and to
account for and disentangle deficiencies in our calculations.  The innovative
use of Bayesian methods will be essential.

\end{itemize}
In short, the framework developed in the last few years to combine LQCD,
EFT, and ab initio nuclear structure is not yet efficient enough to
allow a genuine assessment of uncertainty.  To be of real use in the search for
new physics, all three ingredients must be improved in the coming decade and
made more computationally efficient; their uncertainty also needs to be reliably addressed. But these kinds of intelligent improvements will not, on their own, be enough: increased access to computing resources and dedicated exascale allocations will also be important.

The NSF Project Scoping Workshop that led to this white paper was organized by Jon Engel, Witek Nazarewicz, and Daniel Phillips, and held virtually on January 31 and February 1, 2022. After reviewing the experimental and theoretical status of the field and discussing recent developments, the attendees set forth the challenges to interpreting experimental results, discussed ways to address those challenges, mapped out a path forward, and planned this report. More details about the workshop (list of participants, program, presentations) can be found on its \href{https://a51.lbl.gov/~0nubb/nsf_0nubb/}{website}.

A recent
Snowmass white paper \cite{Snowmass2} provides a particle-physics perspective on these issues.  
The Project Scoping Workshop, and this report, are focused more on the nuclear-theory aspects of  $0\nu\beta\beta$ calculations. The nuclear- and particle-theory for $0\nu \beta \beta$ decay is also reviewed---together with the experimental situation---in Ref.~\cite{AgostinietalRMP}.
Our workshop and this report differ from these papers in the detail in which they discuss modern methods and the accompanying uncertainty quantification that will allow a meaningful error bar in the prediction of {\NME} for any particular BSM mechanism. 

In the next section we provide a summary of the current state-of-the-art in both the nuclear-physics aspects of {\NME} (Sec.~\ref{sec:physics}) and uncertainty quantifcation (UQ) in nuclear theory (Sec.~\ref{sec:UQcontext}). Section~\ref{sec:physicsfuture} then discusses the innovations and calculations that are needed to advance the nuclear theory of {\NME}, while Sec.~\ref{sec:UQ} describes a plan to quantify uncertainty in those calculations. 
Because of the significant emphasis on UQ for $0\nu \beta \beta$ matrix elements in this report Secs.~\ref{sec:UQ} and \ref{sec:UQcontext} are quite detailed and explicit about how we think that UQ can be carried out. 
We close in Sec.~\ref{sec:summary} with a summary of the theory advances and collaborative structures that are needed in order to establish precise and accurate calculations of neutrinoless double-beta decay.

\section{Summary of the current state of the art}

\subsection{Physics}
\label{sec:physics}
Much of the current state of the art in the computation of {\NME} arose from work in the DBD Topical Theory
Collaboration.  LQCD, EFT, and nuclear many-body
methods all played a role in the multi-scale problem.  We discuss recent
developments in each of these areas.

\textsf{\textbf{EFT.}} Before the Topical Collaboration, the connection of nucleon
operators with
fundamental sources of lepton-number violation tended to be \textit{ad hoc},
with BSM models analyzed individually, and unsystematically.  The first
application of 
the framework of chiral EFT to the problem, for $0\nu\beta\beta$ decay
induced by heavy-particle exchange, appeared in Ref.\ \cite{Prezeau03}.  In the
last few years, work of this kind has grown much more systematic.  References\
\cite{Cirigliano17,Cirigliano:2017djv,Cirigliano:2018yza} systematized the work of Ref.\ \cite{Prezeau03}, showing how
the parameters that determine the rates of very heavy-particle lepton-number
violating physics work their way down into nucleon-level $\beta\beta$ 
operators.  At around the same time, Ref.\ 
\cite{Cirigliano:2017tvr} treated
light-neutrino exchange, showing that working to N$^2$LO requires
``non-factorizable'' diagrams (those that cannot be broken in two by cutting the
line representing the exchanged neutrino) that had never been considered before.
Shortly after that, researchers made the surprising discovery
\cite{Cirigliano:2018hja}
that a contact interaction, representing the effects of high
virtual-neutrino momenta that are integrated out of the chiral EFT, occurs at
leading order.  Though the coefficient of the contact operator was initially
unknown, it was later determined approximately through a resonance-model-based
interpolation between perturbative QCD and low-energy pion and nucleon dynamics.~\cite{Cirigliano21a,Cirigliano2021}.  For the first time, nuclear many-body
computations of {\NME} in the nuclei used in experiments are taking the contact
term into account. So far it has caused all ab initio matrix elements to
increase.

\textsf{\textbf{LQCD.}} The hope is that LQCD will soon be able 
to directly supply the
coefficient of the aforementioned contact terms, as well all other relevant LECs.  In the
last few years, the field has made significant progress toward that goal.  A contribution to $0\nu\beta\beta$ decay with TeV-scale LNV is produced by the
exchange of BSM heavy particles between two pions, each
of which are then absorbed by protons as they turn into neutrons.  The exchange
between these virtual pions is easier to compute with LQCD than the direct
exchange between nucleons, and in recent work the dependence of the resulting
$0\nu\beta\beta$ nucleon-level matrix elements on parameters that specify BSM
models has been calculated \cite{Nicholson:2018mwc,Monge-Camacho19}. Pionic matrix elements in the light-neutrino exchange scenario have also been computed in LQCD, and the corresponding LECs in chiral perturbation theory have been constrained~\cite{Tuo:2019bue,Detmold:2020jqv}.  We anticipate progress toward direct calculations of $nn \to pp$ matrix elements and  are developing the formalism for constraining contact LECs from future LQCD calculations~\cite{Briceno:2015tza,Davoudi:2020gxs}.

\textsf{\textbf{Nuclear Structure.}} At the nuclear-structure scale, recent 
progress has been mostly in applying
newly developed non-perturbative ab initio many-body methods to
$\beta \beta$ decay.  Such methods start with interactions and operators
determined from QCD and/or fit to data in very light nuclei ($A=2$ , 3, or 4),
and then produce (approximate) solutions to the Schr\"odinger equation in
heavier nuclei.  Three distinct ab initio methods have been applied together with chiral-EFT interactions to
the heavy open-shell nuclei of interest for $0 \nu \beta \beta$ experiments.   The first two, the In-Medium Generator Coordinate Method (IM-GCM) and the Valence Space IMSRG (VS-IMSRG) are variants of the In-Medium Similarity Renormalization Group (IMSRG), an approach in which one uses renormalization-group flow equations to decouple a predefined ``reference'' state, ensemble, or subspace from the bulk of the many-body Hilbert space.  
The third method is  Coupled Cluster (CC) Theory; it
uses an ansatz for the ground state in which particle-hole excitation operators are exponentiated before being applied to a Slater determinant.

\begin{figure}[t]
    \centering
    \includegraphics[width=0.85\textwidth]{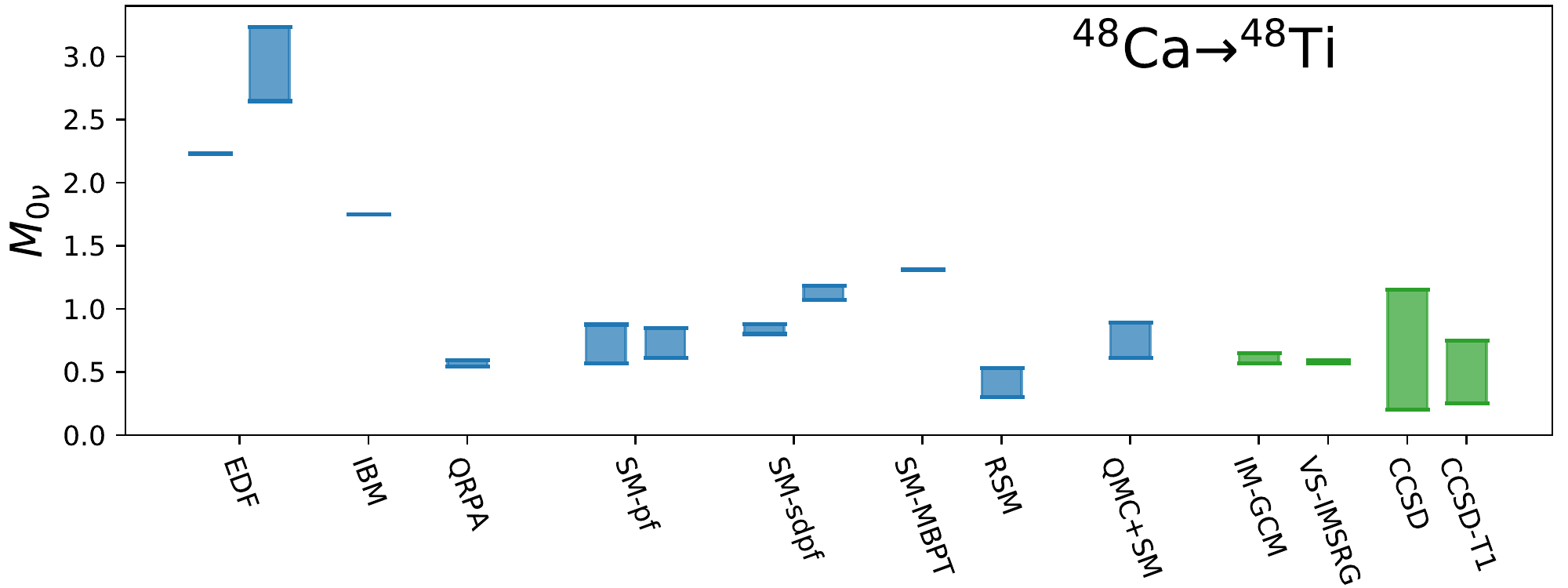}
    \caption{The light-neutrino-exchange \NME\ for the transition $^{48}{\rm Ca}\to^{48}{\rm Ti}$, computed in various approaches. The four right-most values, in green, all result from the the same chiral-EFT interaction. References: EDF~\cite{Vaquero2013,Yao2015}, IBM~\cite{Barea2015}, QRPA~\cite{SimkovicRodin2013}, SM-pf~\cite{Senkov2013,Menendez2009}, SM-sdpf~\cite{Iwata2016}, SM-MBPT~\cite{Kwiatkowski2014}, RSM\cite{Coraggio2020}, QMC+SM~\cite{Weiss2021}, IM-GCM~\cite{Yao2020}, VS-IMSRG~\cite{Belley2021}, CCSD,CCSD-T1~\cite{Novario2021}. }
    \label{fig:Ca48}
\end{figure}

All three of these methods, along with many older and more phenomenological 
schemes, have been applied to the computation of \NME\ for light-neutrino exchange
in $^{48}$Ca, the lightest isotope that can be used in an experiment.  Figure 
\ref{fig:Ca48} displays the compiled results.  Those produced by the methods just described---the ``most ab initio''---are shown in green on the right of the figure.  The uncertainty range is more significant for these than for other methods, but still omits most systematic error.  

The next section reviews the state of the art in uncertainty quantification.  The rest of this document then discusses both the ways in which physics methods can be improved in accuracy, and the ways in which remaining uncertainty in their predictions for \NME\ can be reliably estimated.

\subsection{Uncertainty quantification}

\label{sec:UQcontext}
In 2011 {\it Physical Review A} published an Editorial that stated ``\ldots there is a broad class of papers where estimates of theoretical uncertainties can and should be made. Papers presenting the results of theoretical calculations are expected to include uncertainty estimates for the calculations whenever practicable.''~\cite{PRAEditorial}. Uncertainty quantification is crucial for calculations of $0 \nu \beta \beta$-decay in nuclei. The planning and---eventually, we hope---the interpretation of $0 \nu \beta \beta$-decay measurements requires that theorists deliver not just a expectation value for {\NME}, but also an uncertainty that represents the range of probable values that matrix element can take and does so in a statistically meaningful way. The goal of the uncertainty quantification (UQ) is not a precise evaluation of whatever is missing from the calculation. Quoting again from Ref.~\cite{PRAEditorial}: ``The aim is to estimate the uncertainty, not to state the exact amount of the error or provide a rigorous bound.''

UQ in nuclear-physics calculations pre-dated  that Editorial but standard regression analysis was prevalent for many years~\cite{Dobaczewski2014}. Since then, nuclear-theory UQ has become much more sophisticated. This progress has taken place on several fronts.

The first, and most straightforward kind of UQ, is the estimation of error bars on the parameters $\mathbf{\theta}$ in the nuclear-physics model being employed, then the propagation of those uncertainties---including their correlations---to model predictions. An early example of such an effort is the estimation of the parameters in a sophisticated nuclear energy-density functional~\cite{McDonnell2015}. There are many recent examples of nuclear-structure calculations that do this, but a particularly striking one from the ab initio world constrained the parameters of nuclear forces using data from light nuclei and propagated the resulting uncertainties to predictions for properties of ${}^{208}$Pb~\cite{Hu2021}.

Three pieces of theory technology are commonly employed in such studies:
\begin{itemize}
    \item Bayes' theorem, which relates the multi-dimensional posterior probability density $p$ of the model parameters ${\theta}$ to the data, $y$, used to constrain those parameters, according to:
    \begin{equation}
     p({\theta}|y) \propto p(y|{\theta}) p({\theta}),
    \end{equation}
    where $p({\theta})$ is the {\it a priori} distribution of the parameters $\theta$.

    \item A method by which a representative set of samples of the posterior probability distribution  $p({\theta}|y)$ can be obtained. Markov Chain Monte Carlo (MCMC) sampling is commonly employed, and was combined with a technique called ``history matching'' in Ref.~\cite{Hu2021}. We note that with such a set of samples in hand, it is conceptually straightforward to obtain a predictive probability distribution for, say, {\NME}. That distribution is found by repeated forward evaluation of the model for {\NME} at the different parameter values ${\theta}$ in the set of samples.
    
    \item Emulators that allow rapid evaluation of the (approximate) model at different values of the parameters ${\theta}$. This makes practical the computation of the likelihood $p(y|{\theta})$ within whatever sampling framework is chosen.
\end{itemize}

Nuclear theorists have also become more attuned to the imperfections in their models. The inclusion of a ``model discrepancy'' term in the analysis of data is known to be crucial for accurate parameter estimation~\cite{KoH,BoH}. This means that one must admit that not just  data, but also calculations, have imperfections that may cause a mismatch between theory and experiment. This idea can be formalized as,
\begin{equation}
    y_{\rm exp}(x)=y_{\rm th}(x;{\theta}) + \delta y_{\rm exp}(x) + \delta y_{\rm th}(x),
    \label{eq:KOH}
\end{equation}
where the last two terms encode, respectively, the experimental error (often taken to be independent at the $x$'s corresponding to different data $y$) and the theory uncertainty (which is almost certainly {\it not} independent, i.e., we expect to be correlated across different $x$'s). Significant effort has gone into building models of $\delta y_{\rm th}$ for EFT calculations~\cite{Furnstahl2015,Melendez2019}; since EFT methods are characterized by a systematic expansion in a small parameter, one can predict how they will fail and so write down candidate functional forms for $\delta y_{\rm th}$. But, even when such control is not available, model defects can still be productively introduced, e.g., Gaussian processes can be used to model the discrepancy between density-functional-theory calculations of masses and experimental data thereon~\cite{Neufcourt2018}.

Ultimately, though, the complex dynamics of nuclei means that different theoretical models will be employed to describe them. 
This diversity of models is advantageous because the methods have complementary strengths but also different systematic model discrepancies.
This becomes a virtue by exploiting
the third area of progress, which has been in the use of forms of Bayesian Model Averaging (BMA) or Bayesian Model Mixing (BMM) to incorporate insights from different models into a unified prediction in a statistically rigorous way. BMM can only be done reliably if individual models ${\cal M}_k$ have had their uncertainties quantified in the ways described in the previous two paragraphs. Once that has taken place, the predictions of those models for the observable of interest $y^*$ can be weighted according to ``scoring criteria'':
\begin{equation}\label{BMA}
    p(y^*|y,y_{ev})=\sum_k w_k(y_{ev}) p(y^*|y,{\cal M}_k).
\end{equation}
Here, $p(y^*|y,{\cal M}_k)$ is the posterior for the observable $y^*$, given the data $y$, in a particular model ${\cal M}_k$, and $w_k(y_{ev})$ is a weight that is determined by the model's performance on a  target (or evidence) dataset, $y_{ev}$.  While in the  BMA expression (\ref{BMA}) the weights are constant across the domain, in the more advanced BMM they can also depend on $x$.  We pause here to make two crucial points:
\begin{itemize}
    \item The data, $y_{ev}$ that are used to assess the aspects of model performance that are pertinent for predicting $y^*$ need not be the same as the data set(s) used to calibrate the models. Ideally, the data $y_{ev}$ will be chosen because they are understood to be, or analyzed to be, a good proxy for the quantity of interest, $y^*$, i.e., models' ability to predict $y^*$ is highly correlated with their ability to predict whatever observables are selected to be part of $y_{ev}$.
    
    \item We note that that performance will almost certainly be addressed within the context of the model discrepancy $\delta y_{\rm th}$ of each model, and hence an understanding of those model discrepancies plays a critical role in between-model UQ. 
\end{itemize}
Early nuclear-physics applications of model averaging can be found in Refs.~\cite{Neufcourt2019,JETSCAPE:2020shq}.

In 2020, the Bayesian Analysis of Nuclear Dynamics (BAND) collaboration~\cite{BAND}  began its effort to lower the barrier for nuclear theorists to perform all three of these types of uncertainty quantification. A particular interest within BAND is methodological work on BMM. The main product the collaboration seeks to deliver is software packages and use cases that facilitate emulation, model calibration, and model mixing. The goals of BAND are described in Ref.~\cite{Phillips2021}. 

Within the context of the $0 \nu \beta \beta$ Topical Collaboration, some efforts were proposed to quantify the uncertainties in calculations of ab initio {\NME}. However, these efforts were limited by the ability to rapidly evaluate these matrix elements for the nuclei of interest in $0 \nu \beta \beta$ experiments. This made it difficult to even accomplish the first, parametric, kind of UQ.  The determination of model discrepancy for the different many-body methods employed in $0 \nu \beta \beta$ studies  remains a topic of forefront research, see Sec.~\ref{sec:UQ}.

\section{Physics progress required}
\noindent
Future efforts in the community to deliver reliable $0\nu\beta\beta$ nuclear matrix elements will probably focus on advancing LQCD and EFT calculations of the underlying matrix elements in the few-nucleon sector, ab initio nuclear many-body calculations that use the LQCD and EFT input in experimentally-relevant isotopes. We discuss these subjects in this section and lay out a path for rigorous UQ in the next section. 

\label{sec:physicsfuture}

\subsection{Lattice QCD and Effective Field Theory}
The goal of a combined EFT and LQCD effort in the $0\nu\beta\beta$ program will be the identification and computation of the LECs multiplying 
$nn\rightarrow pp$ transition operators in chiral EFT. 
Because BSM LNV interactions originally involve leptons and quarks, one has to evaluate matrix elements of quark operators in hadronic states in order to link LNV parameters such as $m_{\beta \beta}$ to the LECs.
The program of constructing consistent and predictive nuclear EFTs 
has a long history and a recent review summarizing its status and prospects 
can be found in Ref.~\cite{Hammer2019}.

As mentioned in Sec.~\ref{sec:physics}, 
analyses of the $nn \to pp$ $0\nu \beta \beta$ amplitude revealed that new $nn \to pp$ contact interactions 
are needed at leading order in chiral EFT, even 
for light Majorana-neutrino exchange~\cite{Cirigliano:2017tvr,Cirigliano:2018hja}.
The associated LEC, called $g_\nu^{NN}$, is not determined by symmetry 
considerations or experiment 
(at least not in a straightforward way) and so must be obtained theoretically. 
The LEC $g_\nu^{NN}$ has been studied so far by applying both large-$N_c$ and dispersive methods,
while LQCD studies require methods that are still under development. 
Large-$N_c$ QCD~\cite{Richardson:2021xiu} relates  $g_\nu^{NN}$ to LECs that can
be extracted from the charge-independence-breaking 
combination of nucleon-nucleon scattering lengths in the $^1S_0$ channel. 
Meanwhile, the dispersion-theory approach, inspired by the Cottingham formula for 
electromagnetic hadron masses~\cite{Cirigliano:2020dmx,Cirigliano:2021qko},
leads to a prediction for 
the $nn \to pp$ amplitude near threshold, from which  $g_\nu^{NN}$ can be extracted 
in any EFT regularization and  renormalization scheme, including those
used in nuclear many-body theory. (See Ref.~\cite{Wirth:2021pij} for an 
early use of this $nn \to pp$ input in ab initio nuclear many-body calculations.)
The main uncertainty in this approach comes from inelastic intermediate states  
that can appear  between the two insertions of the weak current (e.g. $NN\pi$ states). 
The existing estimates of  $g_\nu^{NN}$ can be improved by
analyzing suitable $\Delta I =2$ observables, thus anchoring 
$g_\nu^{NN}$ to data. 
Finally, as discussed below, LQCD 
can play a major role in a first-principles determination of  $g_\nu^{NN}$~\cite{Cirigliano:2020yhp,Davoudi:2020gxs, Davoudi:2021noh}. 

The  leading-order LECs associated with TeV-scale LNV operators 
are currently completely unknown. 
Their determination will be possible through the use of dispersion-theory techniques similar to 
those developed in Ref.~\cite{Cirigliano:2020dmx,Cirigliano:2021qko}, as well as by through a direct calculation in LQCD. 
In fact, direct LQCD calculations can in principle determine the entire $nn\rightarrow pp$ amplitude. (See Refs.~\cite{Davoudi:2020ngi, Drischler:2019xuo} for recent reviews of the role of LQCD in constraining nuclear observables).
The interplay between LQCD and EFT is symbiotic: On one hand, matching EFT and LQCD will enable an assessment of the theoretical foundation of nuclear EFT and a calibration of its truncation scheme. 
On the other hand, EFT descriptions 
allow better quantification of the systematic uncertainties in LQCD calculations, providing reasonable extrapolation forms for taking continuum and infinite-volume limits. Furthermore, in order to play a role in the $0\nu\beta\beta$ program,
LQCD calculations need to be performed at quark masses that are sufficiently close to the physical values to allow reliable extrapolations to the physical point. Such extrapolations rely on EFTs, which in turn rely upon LQCD input to determine the relevant LECs. Thus, an interplay between LQCD calculations of two-nucleon ($NN$) observables and EFT will be necessary to determine at which quark masses one may trust results for $0\nu\beta\beta$-decay observables.

Before calculating the $nn\rightarrow pp$ amplitudes with LQCD, however, the low-energy spectra and scattering amplitudes in the $NN$ system need to be calculated with precision. Doing so allows one to determine which operators couple sufficiently well to the ground states of interest, to understand the systematic uncertainties inherent to $NN$ systems, and to match finite-volume Euclidean matrix elements to infinite-volume transition amplitudes.
The $NN$ studies to date have been largely carried out at very large quark masses, where extrapolation to the physical point cannot be controlled. Fully understanding the systematic uncertainties will become even more crucial as the quark masses are lowered toward their physical values because of a signal-to-noise problem for nucleons, in which statistical noise grows exponentially with the pion mass, atomic number, and Euclidean time~\cite{Parisi:1983ae,Lepage:noise,Beane:2009gs}. Furthermore, in calculations at lighter quark masses (which require larger lattice volumes), the energy gaps that dictate the exponential decay of excited states with Euclidean time become very small, causing a slow approach to the ground state that may be obscured by the growth in noise. Thus, improved operators, analysis, and understanding of excited-state contamination are of critical importance. 

These complications mean that we
still do not know whether two nucleons form bound states, even at large quark masses that make precise calculations easier. Recent work within the LQCD community has highlighted the importance of fully-controlled calculations in the $NN$ sector. First, the use of improved interpolating-operator sets and analysis techniques based on the variational principle of quantum mechanics has led to results~\cite{Francis:2018qch,Horz:2020zvv,Amarasinghe:2021lqa} that cast doubt on earlier spectroscopy calculations at similar quark masses. Second, a preliminary study of the discretization effects of two-baryon calculations has shown large shifts in the binding energies
away from the continuum limit~\cite{Green:2021qol}. The latter finding, in particular, needs to be verified by different groups with different lattice actions, and may indicate that $NN$ calculations must be performed at multiple fine lattice spacings.
This would significantly increase the cost of calculations.

To use LQCD to access the $0\nu\beta\beta$ amplitude, one needs to develop indirect mapping relations. This is because the notion of asymptotic states is absent in the finite Euclidean spacetime that is that used in the LQCD setting. 
A general mapping exists to obtain matrix elements of local (short-range) operators such as those appearing in the high-scale models of $0\nu\beta\beta$ decay within two-nucleon states~\cite{Briceno:2015tza}. As an input, this mapping requires two-nucleon spectra and the energy dependence of elastic scattering amplitudes. The existing mapping for the matrix element associated with light neutrino-exchange involves a matching to the leading-order nucleonic EFT~\cite{Davoudi:2020gxs}, and requires as input the two-nucleon spectra and scattering amplitudes. The mappings for such long-range matrix elements are in general more complex than those for local matrix elements because a straightforward analytic continuation in the presence of on-shell intermediate states is not possible~\cite{Briceno:2019opb,Davoudi:2020xdv, Christ:2015pwa, Feng:2020nqj}.  With properly infrared-regulated neutrino propagators, however, analytic continuation will be straightforward in future calculations of the $0\nu\beta\beta$ amplitude~\cite{Tuo:2019bue,Detmold:2020jqv}.
Techniques for computing both the short- and long-distance contributions to $0\nu\beta\beta$ processes have already been developed and applied in studies of the $\pi^-\rightarrow \pi^+ e^- e^-$ and $\pi^-\pi^-\rightarrow e^-e^-$ processes~\cite{Nicholson:2018mwc,Tuo:2019bue,Detmold:2020jqv}, which also constrain pionic contributions within nuclear $0\nu\beta\beta$ decays. Calculations of the $nn\rightarrow pp e^-e^-$ process will be significantly more involved for the reasons discussed above, but will be a crucial next step.

With the broad goal of achieving a systematic quantification of nuclear uncertainties, we must  
face the challenges of extending the analysis of the $0\nu\beta\beta$ transition operator beyond leading order in chiral EFT, especially in the case of 
light-neutrino exchange. We must also go beyond two-nucleon observables to reliably determine the role of multi-nucleon effects in double-beta decay.  Regarding the $nn\rightarrow pp$ amplitude, 
both in Weinberg's power counting (WPC) and in renormalized chiral EFT, the first corrections arise at next-to-next-to-leading order (N$^2$LO) \cite{Cirigliano:2019vdj}. In the two-body sector, the transition operator includes contributions from the nucleon vector, axial, and induced pseudoscalar form factors (which are customarily included in nuclear calculations), from pion-neutrino loops \cite{Cirigliano:2017tvr}, and from new contact
interactions required to absorb the divergences in these loops.
These include the couplings of two electrons to two pions
($g_\nu^{\pi\pi}$), to two nucleons and one pion $(g^{\pi N}_\nu)$, and to four nucleons (a correction to $g_\nu^{NN}$). $g_\nu^{\pi\pi}$ is well determined by LQCD~\cite{Tuo:2019bue,Detmold:2020jqv}, while extracting the correction to $g_{\nu}^{NN}$
will require the matching of LQCD and chiral-EFT amplitudes at higher orders.  The short-range structure of the two-body $0\nu\beta\beta$ operator at N$^2$LO is at the moment unknown beyond WPC. Reference~\cite{Cirigliano:2019vdj} pointed out that
the promotion of $g_\nu^{NN}$ to LO implies that certain derivative operators in the spin-singlet channel are also more important than in WPC, but a full analysis of the
LNV scattering amplitudes to N$^2$LO does not yet exist,
and needs to be developed to interpret anticipated LQCD results. A deeper question is whether the chiral and momentum expansions of chiral EFT converge (and converge to what is observed in Nature). This question is open even for
single baryons~\cite{Chang:2018uxx,Drischler:2019xuo}
and relatively light systems~\cite{Yang:2020pgi,Yang:2021vxa,Tews:2022yfb},
and needs to be answered as the community moves beyond purely phenomenological approaches. LQCD input for the unknown LECs at successively higher orders can help resolve power-counting questions for LNV processes.

Moving to the question of multi-nucleon corrections, we note that 
two-body currents, which are important in the $g_A$-quenching problem in $\beta$ decays~\cite{Towner:1987zz,Gysbers:2019uyb}, first contribute to $0\nu\beta\beta$ decay at N$^2$LO, by generating  three-body corrections to the operator. These corrections were considered in
Refs.~\cite{Menendez:2011qq,Wang:2018htk} and found to be compatible with power-counting estimates.
Furthermore, in the three-body sector, a goal for the chiral-EFT community is to validate or falsify WPC's expectations, by studying suitable few-body amplitudes. 
Once calculations of two-body transitions have been achieved with systematic control, LQCD studies of $0\nu\beta\beta$ decay of $A\in\{4,6\}$ systems, if they can be carried out, will provide valuable additional information. Such calculations can reduce systematic uncertainties in the process of matching $0\nu\beta\beta$ amplitudes to the chiral EFT used in nuclear many-body calculations. In particular, constraining the same LECs from LQCD calculations of different processes will not only reduce statistical uncertainty, but will also, through benchmarking, reduce the uncertainties in nuclear EFTs that arise from choices of scheme or regulator. Useful transitions will probably include the $A=4$ processes
$^4\text{H} \rightarrow {^4\text{Li}}\, e^-e^-$
and $^4\text{n} \rightarrow {^4\text{He}}\, e^-e^-$,
and the $A=6$ transitions $^6\text{He}\rightarrow {^6\text{Be}}\, e^-e^-$
(for which nuclear-structure calculations have been performed~\cite{Pastore:2017ofx}),
and $^6\text{H} \rightarrow {^6\text{Li}}\, e^-e^-$
(which introduces additional challenges for many-body approaches 
because $^6\text{H}$ is unstable). To reduce the cost of extrapolating such LQCD calculations to infinite volume, directly matching matrix elements to finite-volume EFT calculations~\cite{Barnea:2013uqa,Kirscher:2017fqc,Detmold:2021oro,Detmold:2004qn} may be a valuable approach to precisely determining the LECs.

In summary, while significant outstanding challenges must be overcome to reliably determine the $nn\rightarrow pp$ amplitude, for both the light-neutrino exchange and the short-distance $\Delta I=2$ 4-quark operators, there exists a clear road map for addressing them.  Following it will require a concerted effort in LQCD, EFT, and the coupling of these theories, as well as computing resources at the exascale and beyond, both to quantify the uncertainties in the relevant two-body process and to build an understanding of multi-nucleon corrections.  

\subsection{Many-body methods}

All experimentally relevant $0\nu\beta\beta$ candidate nuclei, with the exception of $^{48}$Ca, are open-shell and at least of medium mass.
Consequently, only a subset of the currently-available ab initio many-body methods can be used to compute the nuclear matrix elements that govern their decay. First computations of the nuclear matrix elements have been performed in coupled cluster theory~\cite{Novario2021}, the IM-GCM~\cite{Yao2020}, and the VS-IMSRG~\cite{Belley2021}.  We describe each of these methods and prospects for improving them next.

\subsubsection{Coupled cluster method}
In the coupled cluster method~\cite{kuemmel1978,bartlett2007,Hagen2014}, the exact wave function $|\Psi\rangle$ is parameterized by the exponential ansatz $|\Psi\rangle =e^{\hat{T}}|\Phi_0\rangle$, where the reference $|\Phi_0\rangle$ is a product state, and the cluster operator $\hat{T}$ generates particle-hole excitations.
One expresses $\hat{T}$ in terms of single, double, triple etc.\ particle-hole excitations and (usually) truncates it at the so-called doubles or triples level. This is the main approximation.
The calculation of transition matrix elements in coupled cluster theory is complicated by the fact that $\hat{T}$ is purely an excitation operator, i.e.\ the fact that the similarity transform $e^{-T}\hat{O}e^T$ of a Hermitian operator $\hat{O}$ is not Hermitian. This implies that the bra version of a state needs to be parameterized through a de-excitation operator rather than an excitation operator.
An additional complication for $0\nu\beta\beta$ decay is that the initial and final states are the ground states of different nuclei, with each in principle requiring its own $\hat{T}$ operator and reference state.
In practice, one expresses the final state as a generalized excitation of the initial state through the equation-of-motion method as $|\Psi_F\rangle = e^{\hat{T}}\hat{R}|\Phi_0\rangle$, where $\hat{R}$ is a double-charge-changing excitation operator~\cite{Novario2021}. Alternatively, one can express the initial state as an excitation of the final state.
In the absence of any truncation, these two choices should yield identical results, so the difference between the two is an indication of the truncation error.

Like $\hat{T}$, the excitation operator $\hat{R}$ is expanded in terms of charge-changing few-nucleon ``excitations'' and truncated at a doubles or triples level. This approximation may not be accurate when the initial and final nuclei are very different in structure, because, for example, they differ in their intrinsic deformation. 
Indeed, the spherically-symmetric coupled cluster method works well for computing properties of closed-shell nuclei such as $^{48}$Ca.
However, the ground state of $^{48}$Ti (the final state in the decay of $^{48}$Ca) is open-shell and is better treated with an intrinsically deformed, (though axially symmetric) reference state, which is computationally more expensive.
In benchmarks performed so far~\cite{Novario2021}, it appears that taking $|\Phi_0\rangle$ to be the deformed $^{48}$Ti state yields more accurate results, though the reason is not entirely known. 

We can expect this approach to be applied to more nuclei and with more accuracy in the next few years.  With enough computation time, it can be generalized to allow triaxial deformation of the reference state (and thus a good calculation, e.g., in $^{76}$Ge) and the restoration of rotational symmetry through projection onto states with good total angular momentum~\cite{Hagen2022}.  These developments will turn the method into a much more versatile tool for the computation of $0\nu\beta\beta$ nuclear matrix elements.

\subsubsection{IM-GCM}
The IM-GCM~\cite{Yao2018,Yao2020} is a combination of the GCM \cite{Reinhard1987} and the Multi-Reference In-Medium Similarity Renormalization Group (MR-IMSRG)~\cite{Hergert:2017kx,Hergert2018}. 
(``Multireference'' refers to a generalization of the renormalization-group flow equations to work with a reference state that is more complex than a Slater determinant.)
The GCM efficiently captures the collective long-range correlations which are important in deformed nuclei, while the MR-IMSRG captures short-range correlations associated with the repulsive core of realistic nuclear interactions.

On can view the IMSRG as a way to generate a unitary transformation $U$ of the Hamiltonian that brings it to a form more amenable to solution.
The transformation is parameterized by a flow parameter $s$, yielding a differential equation for the transformed Hamiltonian $H(s)=U(s) H U^{\dagger}(s)$ and for other consistently-transformed operators $\mathcal{O}(s)=U(s) \mathcal{O} U^{\dagger}(s)$.
The unitary transformation, which is conveniently expressed in the Magnus formulation as $U(s)=e^{\Omega(s)}$~\cite{Morris2015}, is designed so that with increasing $s$, a reference state $|\Phi_0\rangle$ increasingly approximates an eigenstate of $H(s)$.

In the IM-GCM, the approach is to take $|\Phi_0\rangle$ to be the ground state of a GCM calculation.
The GCM ground state is expressed as a linear combination of configurations $|\Phi(q)\rangle$ labeled by a set of generator coordinates $q$ (e.g. quadrupole deformation), so that $|\Phi_0\rangle = \sum_q f(q) |\Phi(q)\rangle$.
The amplitudes $f(q)$ are obtained by minimizing the energy via the Hill-Wheeler-Griffin equation, which amounts to a diagonalization the space
of GCM states $|\Phi(q)\rangle$.  As we noted in the context of coupled cluster theory, the initial and final states in any $0\nu\beta\beta$ decay are different, a fact that complicates most computations.
The transformations $e^{\Omega_I(s)}$ and $e^{\Omega_F(s)}$ that decouple the initial and final states are not equivalent.

In Ref.~\cite{Yao2018}, this complication was addressed by combining the IMSRG transformation and GCM calculations for the initial and final states in different ways, again with the understanding that, without approximations, all these combinations should give the same results. In more recent work, a powerful alternative was presented in the form of an ensemble composed of reference states in both the initial and final nuclei that allows one to use a single transformation rather than two separate ones \cite{Yao2020}.


The main approximation in the MR-IMSRG flow equations is that all operators are truncated at the normal ordered two-body (NO2B) level.  In the next few years, with enough computational capacity, we will be able to go beyond this approximation by either exactly or approximately including the effects of three-body operators that are induced by the flow equations.
A first step in this direction indicated that the correction due to induced three body operators is sub-leading (on the order of 10\% of the NO2B correction)~\cite{Yao2020,Yao:2021bs}.  The result is encouraging, but the corrections are large enough that they should be included.

Another approximation is in the selection of generator coordinates. 
In principle, one can continue to add more coordinates that are believed to be relevant (for example, proton-neutron pairing gaps) and confirm that the answer does not change, but it is difficult to establish that all important degrees of freedom have been explored. Historically, this has been a significant issue for the GCM calculations based on phenomenological interactions. In the IM-GCM, this issue can be overcome because dependence of the transformation on the flow parameter $s$ offers a powerful diagnostic tool: If sufficient degrees of freedom are included in the MR-IMSRG flow and the GCM basis, the unitarity of the transformation will not be spoiled by truncation errors, and all observables should be independent of  $s$   \cite{Gebrerufael:2017fk,Frosini:2021ddm}.

\subsubsection{VS-IMSRG}
In the VS-IMSRG~\cite{Stroberg2019}, as in the IM-GCM, the strategy is to perform a unitary transformation to bring the Hamiltonian into a form more amenable to solution.
In this case, the transformation block-diagonalizes the Hamiltonian such that an additional diagonalization in a valence shell-model space yields exact results (assuming no truncations are made in the transformation).

The VS-IMSRG calculations carried out thus far have generally performed the normal ordering with respect to a closed-shell reference state or an uncorrelated ``ensemble'' reference that has the correct number of protons and neutrons on average.
As with coupled cluster theory, one needs to choose the initial or final state as the reference and, in the absence of truncation, this should not affect the answer.
The simpler reference used in the VS-IMSRG (compared to, e.g., the IM-GCM) is somewhat compensated for by the subsequent exact diagonalization in the valence space, resulting in a complementary approximation scheme.
Like IM-GCM, the VS-IMSRG as currently practiced truncates operators at the two-body level after normal ordering, and the clear next step is the approximate inclusion of the effects of induced three-body operators.  Again, with sufficient computational resources and personpower, this can be done.

\subsubsection{Benchmarking with quasi-exact methods}
Quasi-exact ab initio methods, namely quantum Monte-Carlo (QMC) and the no-core shell model (NCSM), are generally limited to light systems, which are not directly relevant for $0\nu\beta\beta$ experimental searches.
They play an important role, however, in benchmarking the methods we've discussed, which can reach the relevant heavier systems.
The three methods described above, coupled cluster theory, IM-GCM, and VS-IMSRG, have all been benchmarked against the NCSM in light systems up to $^{14}$C (and up to $^{22}$O with the importance-truncated NCSM)~\cite{Basili:2020kq,Yao2020,Yao:2021bs,Novario2021}.
The benchmarks showed that coupled cluster calculations that use a deformed reference state are usually more accurate than those that use a spherical reference.

In contrast to the NCSM, IM-GCM, VS-IMSR, and CC theory, QMC approaches do not rely on a single-particle basis expansion. Variational Monte Carlo (VMC) approximates the solution of the many-body problem by an accurate trial wave function $\Psi_T$, obtained by applying two- and three-body correlation operators to a Slater determinant of $A$ single-particle wave functions~\cite{Carlson:2014vla,Gandolfi:2020pbj}. The optimal set of variational parameters defining the trial wave function is obtained by minimizing the energy expectation value $\langle \Psi_T|H|\Psi_T\rangle$ with dedicated optimization algorithms~\cite{Contessi:2017rww}. The limitations of the variational ansatz are overcome by the Green's function Monte Carlo (GFMC) method that propagates the trial wave function in imaginary-time to extract the ground-state of the system $|\Psi_0\rangle = \lim_{\tau\to\infty} |\Psi(\tau)\rangle = \lim_{\tau\to\infty}{\rm exp}[-(H-E_0)\,\tau]\,|\Psi_T\rangle$. QMC methods have no difficulty in using ``stiff'' forces that can generate wave functions with high-momentum components, but they are limited to local (or nearly local) Hamiltonians because non-localities exacerbate the fermion-sign problem~\cite{Lynn:2012fq}. There have been QMC studies of the $0 \nu \beta \beta$-decay nuclear matrix elements for light  nuclei (see, e.g., \cite{Pastore:2018ph,Wang:2019qz}), but the (nearly) local Hamiltonians \cite{Gezerlis:2014zr,Lynn:2016ec,Lonardoni2017,Piarulli:2018xi,Piarulli:2020df} used in these studies pose a substantial hurdle for direct benchmarks against the configuration-space methods that we discussed above.  More recently developed local chiral interactions with typical cutoffs around $\Lambda=500\,$MeV lead to much slower convergence than their nonlocal counterparts with the same scales. Renormalization group transformations may help to mitigate this problem, but the uncertainties due to the omission of induced contributions to the interaction and transition operators may also be more substantial than in a nonlocal regularization scheme. 
Nevertheless, once RG and EFT truncation errors have been propagated to the $0\nu\beta\beta$ matrix element, a comparison of QMC and configuration-space methods will represent an important check.

\subsubsection{Other ab initio approaches}
We have focused on the several ab initio methods that have already been applied to experimentally relevant transitions, but there are others that may also be able to tackle these nuclei soon.
Applications of QMC have been limited almost entirely to the $p$-shell and below because the number of spin/isospin states scale exponentially with particle number $A$ (see, e.g., \cite{Gandolfi:2020fv} and references therein). However, within the auxiliary field diffusion Monte Carlo (AFDMC) approach~\cite{Schmidt:1999lik} the spin-isospin degrees of freedom are described by single-particle spinors, the amplitudes of which are sampled with Monte Carlo techniques based on the Hubbard-Stratonovich transformation.  The transformation reduces the computational scaling from exponential to polynomial in $A$. AFDMC calculations for $^{16}$O have been reported~\cite{Lonardoni2017}, and calculations of $^{48}$Ca are conceivable in the near future. 
 
A recently proposed alternative is to use QMC to compute \NME\ for light nuclei, and match an effective shell-model operator to these calculations~\cite{Weiss2021}, using the generalized contact formalism (GCF). The effective operator is then employed in shell-model calculations of heavier nuclei, where QMC is not feasible. This approach can be viewed as the QMC providing synthetic data to which a shell-model effective operator can be fit. It is justified by the factorization of physics at the scale of nucleon-nucleon interactions from the nuclear environment. This factorization is seen in the application of renormalization group (RG) methods to QMC wave functions, where short-distance physics in those wave functions evolves into effective operators at the lower resolution appropriate to the shell model ~\cite{Tropiano:2021qgf}. The GCF implements the leading-order consequences of factorization. 
One challenge for the future will be quantifying the long-range correlations missed by the shell model; these will in general depend on the valence space (see effective charges for $E2$ transitions as an example). Such quantification is one aspect of a more general question about what the sub-leading corrections to the GCF calculation carried out in Ref.~\cite{Weiss2021} are. 

The RG approach to this problem makes it clear that, in {\it any} of the approaches to the nuclear many-body problem described here, the $0 \nu \beta \beta$ operator must be evolved consistently with the methods used to reduce the effective size of the space in which the many-body wave functions are computed. It follows that the $0 \nu \beta \beta$ contact operator will not necessarily be the same as the one computed in Ref.~\cite{Cirigliano:2021qko}, or obtained in the future from LQCD. Instead that short-distance operator must absorb the physics between the hadronic scale of LQCD/sum-rule calculations and the low-energy nuclear-structure scale; i.e., it will account not just for hadronic excitations that have been integrated out of the Hilbert space, but for high-energy nuclear correlations that are integrated out too. 

The NCSM may also be applied to heavier istopes in the future.  Although the method in its original form is typically limited to $A\lesssim 16$, the importance-truncated NCSM (IT-NCSM) \cite{Roth:2009eu} can significantly reduce the dimensions of the Hamiltonian and thus reach higher in mass, conceivably up to $^{48}$Ca.  One challenge will be to obtain a better understanding of the extrapolation of \NME\ in the importance truncation parameter $\kappa_{\rm min}$.

The symmetry-adapted no-core shell model (SA-NCSM) is a version of the NCSM that uses irreducible representations of the symplectic symmetry group $Sp(3,\mathbb{R})$ rather than particle-hole energy to truncate its basis \cite{Dytrych:2020db,Launey:2020um,Launey:2021cq,McCoy:2020ol}. 
The alternative truncation scheme allows it to efficiently capture deformation, which is important in either the mother or daughter nucleus in all experimentally relevant $0 \nu \beta \beta$-decay candidates. Applications of the SA-NCSM have mostly focused on $p$- and $sd-$shell nuclei so far, but first results for 
$^{48}$Ti
have been reported in Ref.~\cite{Launey:2021cq}. These particular results serve as a demonstration that convergence of a SA-NCSM calculation is mainly affected by the strength of the mixing between irreps in a particular nucleus rather than the mass number: the model space dimension for 
$^{48}$Ti is more than an order of magnitude smaller than the dimension for
$^{20}$Ne.
At present three-nucleon forces have not yet been included in the SA-NCSM, but
once this challenge is overcome, the SA-NCSM will be a valuable complementary approach to the coupled cluster and VS-IMSRG methods that employ particle-hole based truncations. It is also complementary to the IM-GCM because the $Sp(3,\mathbb{R})$ irreps offer a more systematic approach to basis construction than the selection of relevant generator coordinates.

Both the conventional NCSM and SA-NCSM can also be combined with an (MR-)IMSRG preprocessing of the Hamiltonian and transition operators to accelerate convergence \cite{Gebrerufael:2017fk}. A combination of MR-IMSRG evolution and SA-NCSM, in particular, would embrace a similar philosophy as does the IM-GCM.

\subsubsection{Other methods}\label{lowerres}
Besides the ab initio methods described in the previous subsections, a variety of other methods have been used to compute \NME\ in nuclei of interest to experimentalists.  These others can be broadly grouped into categories: the interacting shell model~\cite{Caurier1990,Caurier1996, Caurier2005, HoroiStoica2010, HoroiBrown2013, Horoi2013, NeacsuHoroi2015, NeacsuHoroi2016,18ho035502}, energy-density-functional (EDF) methods~\cite{Rodriguez2010}, the quasi-particle random phase approximation (QRPA) ~\cite{SuhonenCivitarese1998, Simkovic1999, Rodin2006, KortelainenSuhonnen2007, SimkovicRodin2013}, and the interacting boson model (IBM)~\cite{Barea2009, Barea2013}. 

In this report the emphasis is on methods that can, in principle, quantify the theoretical uncertainties of the underlying strong-interaction Hamiltonian and of the transition operators. However, methods such as the ``phenomenological'' shell model still have a valuable role to play, because they preserve underlying nuclear many-body symmetries and thus capture the most relevant degrees of freedom. In addition, the heavy work of finding optimized effective shell-model Hamiltonians and effective transition operators that capture the landscape of realistic nuclear spectra and of the experimentally accessible nuclear observables has already been done in these approaches. We envision that semi-phenomenological methods, such as the shell model, can be used to explore correlations between observables, helping us identify the quantities that best reflect the accuracy of an ab initio calculation. For example, an ensemble of shell model Hamiltonians can be generated by adding random contributions to the two-body matrix elements of some ``seed'' Hamiltonians~\cite{PhysRevC.101.054308}. These Hamiltonians can then be used to obtain {\NME}, as well as excitation spectra and electroweak transitions and moments (for which data exist or could be obtained). Any observables which are significantly correlated with {\NME} would then be explored in the more expensive ab initio calculations, producing input for subsequent model-mixing analysis. An initial study along these lines for the $0\nu\beta\beta$ decay of $^{48}$Ca-$^{48}$Sc-$^{48}$Ti system can be found in Ref.~\cite{Horoi48CaStat22}.

\section{A program for uncertainty quantification}\label{sec:UQ}

\begin{figure}[htb]
    \includegraphics[width=0.7\textwidth]{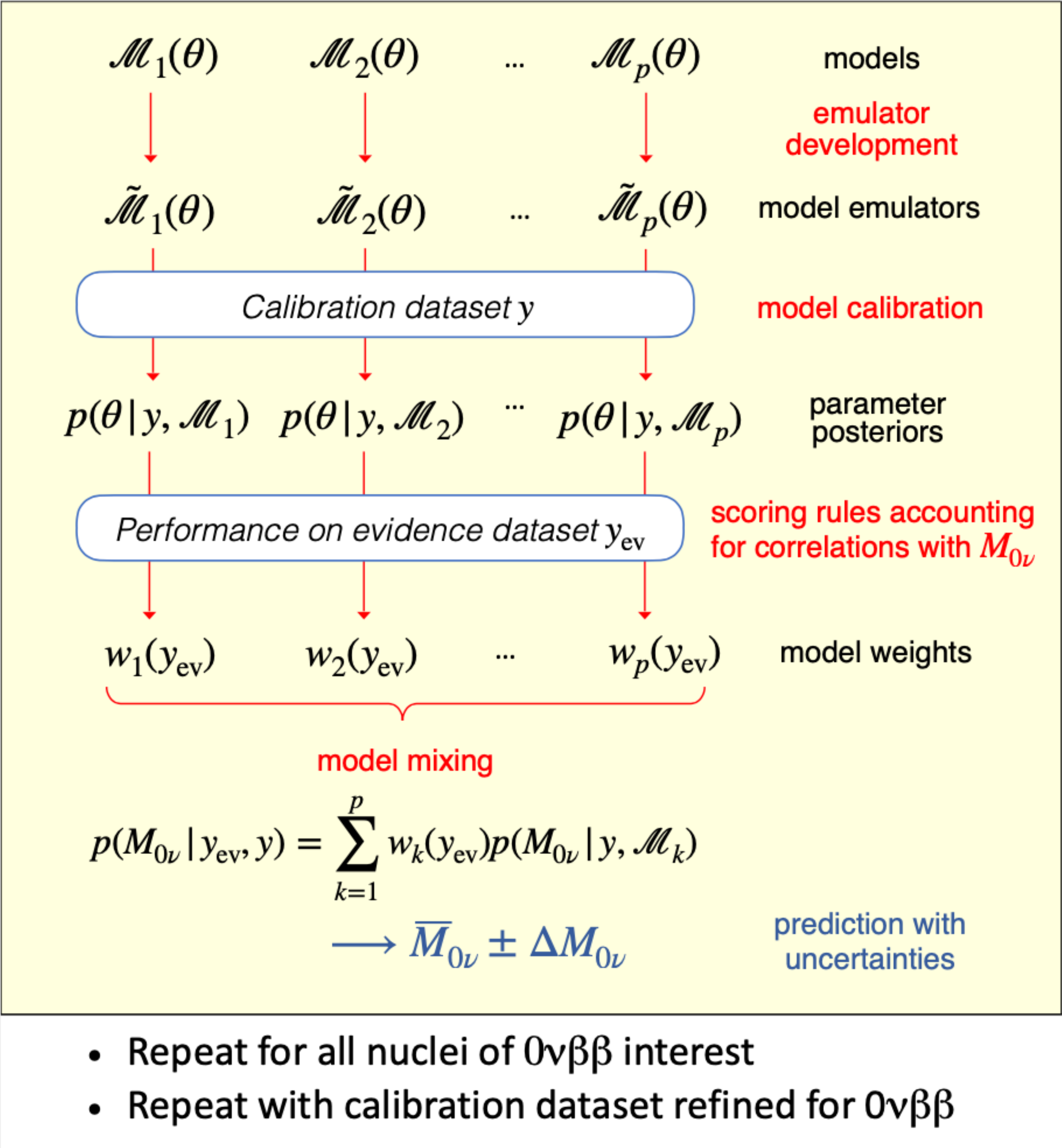}
    \caption{The road to calculations of {\NME} with  UQ that accounts for all limitations of the nuclear-physics calculation: truncation errors in chiral EFT, uncertainties in the theory's parameters, and deficiencies of the many-body methods used. Emulator development is the first step, as it is key to facilitating subsequent calculations.  Model parameters $\theta$ can be calibrated against a dataset $y$. Weights for the different many-body methods will  be obtained by assessing methods' performance on a set of observables $\{y_{\rm ev}\}$. The weights $w_k(y_{\rm ev})$ are to be computed via scoring rules that gauge predictive accuracy~\cite{Yao:2018abc} and will also take account of the extent to which different members of $\{y_{\rm ev}\}$ are correlated with {\NME}.}
    \label{fig:UQ-DBD}
\end{figure}

\noindent
The preceding sections outlined a variety of many-body methods that can be used to perform ab initio calculations of $0\nu\beta\beta$ nuclear matrix elements. 
It might be supposed that the goal of a UQ analysis should be to determine the ``best'' of these methods and that whichever method turns out to be ``best'' should then be used exclusively.
In fact, these methods have complementary strengths and deficiencies, so the goal instead is to use all of them to optimize the overall predictions.
Therefore in this section we outline a procedure, depicted in Fig.~\ref{fig:UQ-DBD}, by which the results for {\NME} obtained in those different methods---as well as their uncertainties---can be combined into a single, unified prediction for {\NME}. We also explain how that procedure will naturally suggest alternative strategies for calibration of the ab initio calculations which should, in turn, lead to refined predictions for {\NME}.

Throughout this section we have in mind that we are considering ab initio predictions for {\NME} that are obtained with chiral-EFT forces and decay operators. Differences between different implementations of the chiral-EFT force should therefore be encompassed within the uncertainty assigned due to truncation of the EFT expansion, cf.\ Sec.~\ref{sec:physics} above. The source of uncertainty that is hardest to assess is therefore that due to the use of different methods for solving the $A$-body problem: these are associated with different ways of truncating the $A$-body Hilbert space. In what follows we denote the different many-body methods that have been, or may in the future be, adopted to solve this problem as $\mathcal{M}_k$. We treat these as different ``models'' in the statistical sense of the term ``model'' and seek to combine their predictions into a single prediction that accurately assesses uncertainties in the evaluation of {\NME}. We assume that uncertainties due to the truncation of the chiral-EFT expansion are reflected in the posterior distribution that must be provided by each many-body method, $\mathcal{M}_k$.

Method $\mathcal{M}_k$'s prediction also  has an inherent
parametric uncertainty, coming both from the parameters of the Hamiltonian used to obtain the wave function of the initial and final state in the double-beta-decay process, and from the contact piece of the $0 \nu \beta \beta$ operator.
In what follows we denote the low-energy constant that multiplies the contact piece by $\eta$ and the parameters of the Hamiltonian as $\theta$.

Recently, $\eta$ has been determined \cite{Wirth2021} by  reproducing the synthetic datum, $y_{\rm synth}$ provided in 
Refs.~\cite{Cirigliano2021,Cirigliano21a}. Meanwhile, for most of this section we will assume that the parameters $\theta$ are calibrated to a dataset $y$ (cf. Sec.~\ref{sec:UQcontext}) that does not have to include  observables that we expect are correlated with $0 \nu \beta \beta$ decay. This is, after all, the stated orientation of most ab initio approaches, which calibrate the parameters of $NN$ and three-nucleon forces (3NFs) to $NN$ scattering data and a few observables in light nuclei. The posterior probability distribution for the parameters $\theta$ that is obtained from such an analysis is denoted $p(\theta|y)$.

From a Bayesian perspective, each many-body method's prediction of {\NME} also comes with a systematic error  that depends on parameters of the approach employed, e.g., Hilbert-space size, accuracy of treatment of 3NFs, etc. The statistical modeling of this systematic error is referred to as discrepancy learning (cf. Eq.~(\ref{eq:KOH})).
Simultaneous learning of discrepancy and parameters is a complicated practical and theoretical exercise. 
Moreover, with no information to leverage near the quantity of interest {\NME}, verification of discrepancy can be difficult.
Nonetheless, grounded, informed priors on the discrepancy can improve prediction---especially when we seek to leverage the predictions made across several many-body methods. We therefore write:
\begin{equation}
   M_{0 \nu}({\rm true})=M_{0 \nu}^{\cal M}(\theta,\eta;\lambda) + \delta M_{0 \nu}^{\cal M}(\lambda),
\end{equation}
where $M_{0 \nu}^{\cal M}(\theta,\eta;\lambda)$ is the prediction obtained in method ${\cal M}$ with method hyperparameters $\lambda$ (and at specific Hamiltonian and operator parameter values) and $\delta M_{0 \nu}^{\cal M}(\lambda)$ is the corresponding model uncertainty. 

If we, for the moment, ignore the issue of the discrepancy function, then the method-${\cal M}$ prediction of {\NME}  is formed by marginalizing over $\theta$ and $\eta$ using the distributions established for them from the data $\{y,y_{\rm synth}\}$:
\begin{equation}
    p(M_{0 \nu} | \mathcal{M}) = \int p(M_{0 \nu}| \theta,\mathcal{M}) p(\theta|y,\mathcal{M}) p(\eta|y_{\rm synth}, \mathcal{M}) \mathrm{d}\theta \mathrm{d} \eta.
\end{equation}
Here it should be noted that we have allowed for the possibility that the probability density obtained for both the Hamiltonian parameters and $\eta$ is different for different methods, i.e., depends on the method ${\cal M}$. We have, however, assumed that all  all methods are calibrated using a common data set $y$. 


But the problem of model discrepancy is critical in predictions for neutrinoless double-beta decay:  different approaches to the nuclear many-body problem are based on different physics assumptions, and so have different model discrepancies. In order to get a handle on the model discrepancy  we propose to assess that method's ability to predict observables that may share similar physics features to the $0 \nu \beta \beta$ decay in the nucleus of interest. Candidate processes include:
\begin{itemize}
    \item Single $\beta$-decay rates in neighboring nuclei, e.g., in the intermediate nucleus in $0\nu\beta\beta$ decay;
    \item $\beta$-strength distributions;
    \item Known $2\nu\beta\beta$ decay rates;
    \item Magnetic moments and $B(M1)$ rates  in three nuclei involved in $0\nu\beta\beta$ decay;
    \item Energies of the lowest $J^\pi=2^+$ states and $B(E2,2^+\rightarrow 0^+)$ rates in initial and final nuclei;
    \item Charge radii;
    \item Observables probing a 100\,MeV momentum-transfer scale, e.g., in muon capture.
\end{itemize}
The idea is then that a method that performs well on these observables, which we denote collectively as $y_{\rm ev}$, should be a more accurate predictor of {\NME} than one that does not.

But which of these observables are most important for constraining the $0\nu\beta\beta$ decay rates? Until now discussion on this point has been largely driven by qualitiative arguments. We propose that, by using properly calibrated Hamiltonians, this question can be answered by analyzing the correlations between the observables on the list above  and 
{\NME}. Those correlations can be well approximated by
 drawing a finite number of samples from $p(\theta|y)$ (say $\approx 100$), using model ${\cal M}$ to compute each observable in the set $y_{ev}$ and {\NME}, and extracting the empirical correlation coefficient of $M_{0 \nu}$ and each quantity in $y_{ev}$ for that model ${\cal M}$.  Examples of such sensitivity studies can be found in, e.g.,  Refs.~\cite{UNEDF0,Reinhard2010,Ekstrom2019}. 
 
 A few supplementary points regarding this correlation analysis need to be made:
 \begin{itemize}
     \item This correlation need not be the same in every method. Different methods have different discrepancy functions, because different methods truncate the nuclear many-body problem in different ways. A particular example of this is that methods with different resolution scales may differ in whether their discrepancy function reflects errors in the long-distance physics or errors in the short-distance physics. Indeed, the balance between these two types of errors could  shift within a particular method as the value of the hyperparameters $\lambda$ changes. It follows that the correlation found for  method ${\cal M}$ at  particular values of $\lambda$ may depend on either $\lambda$ or ${\cal M}$. But, analysis of these correlations, when combined with data on the observables in the set $y_{ev}$, will help us pin down the discrepancy functions, or at least minimize their impact on the {\NME} prediction. 
     
     \item The prediction of the observables $y_{ev}$ may depend on additional parameters $\gamma$, that are not part of the set $\theta$, and are not \emph{a priori} needed to predict {\NME}. The posterior prediction for $y_{ev}$ is then formed by marginalizing over $\gamma$, using a probability distribution function that can be thought of as a prior for our purposes, but may be informed by studies of the pertinent observable(s) in nuclei that are some distance from the $0 \nu \beta \beta$ candidate 
     \begin{equation}\label{pM}
    p(y_{ev} |y, \mathcal{M}) = \int p(y_{ev}, \theta,\gamma, \mathcal{M}) p(\theta|y)p(\gamma|\mathcal{M}) \mathrm{d}\theta\mathrm{d}\gamma,
\end{equation}
Marginalizing over $\eta$ will certainly be necessary for the prediction of {\NME}. Such marginalization (over $\eta$ and $\gamma$) may affect the correlation.

\item As discussed in Sec.~\ref{lowerres}, 
the correlation analysis does not need to be initiated within computationally-expensive ab initio calculations. In the first instance, it can be carried out using  lower resolution approaches that can be viewed as computationally less expensive emulators of ab initio methods. An example of ongoing work along these lines can be found in Ref.~\cite{Horoi48CaStat22}.
 \end{itemize}

Such a correlation analysis is useful in its own right. But, with these correlation coefficients in hand we can form an ``improved Bayesian Model Average'' of the results from the different many-body methods ${\cal M}$. 
In BMA \cite{Phillips2021}, a set of candidate methods $\mathcal{M}_1,\ldots,\mathcal{M}_p$ are combined to form a predictive distribution via Eq.~(\ref{BMA}):
\begin{equation}
    p(M_{0 \nu} | y_{ev},y) = \sum_{k=1}^p w_k(y_{ev}) p(M_{0 \nu} |y, \mathcal{M}_k),
    \label{eq:BMA}
\end{equation} 
where    $w_k(y_{ev})$ represents the underlying weight given to each method.
The weight $w_k$ is proportional to $p(y_{ev}|\mathcal{M}_k)\times p(\mathcal{M}_k)$, where $p(y_{ev}|\mathcal{M}_k)$ represents the evidence for method $\mathcal{M}_k$ present in data $y_{ev}$ and $p(\mathcal{M}_k)$ is a prior probability that method $k$ is correct. This prior is usually taken to be flat across the methods $\mathcal{M}_k$, i.e., the methods are all taken to be equally plausible.
The formula (\ref{eq:BMA}) is aspirational in that there are complications in its deployment that are both practical and theoretical.

The selection of the weights requires careful documentation of the source of the systematic errors between method predictions and observables.
Extraneous observables, i.e., observables that are no more than weakly related to $0 \nu \beta \beta$, are less dangerous to the resulting inference if all error sources were only experimental and independent throughout the observable space as no method has a specific advantageous bias.
However, in the context of $0 \nu \beta \beta$ one expects that a significant portion of the error can be attributed to systematic method deficiencies $\delta y_{\rm th}$.
The result is significantly related error that exists across the observable space.

It is therefore critical to carefully select observables in $y_{ev}$.
Observables that are closely related to the target observable, $y^*=M_{0 \nu}$ should receive higher weights---something that classical BMA does not do. If observables in $y_{ev}$ that are not singificantly related to {\NME} can influence the weight of a method in the BMA formula
(\ref{eq:BMA}), they are likely to dilute---or even bias---the prediction for {\NME}.
Parsimony is also important for a practical reason: the Bayesian model averaging formula requires the same $y_{ev}$ should be used across all many-body methods, meaning all of them need to be able to produce predictions of these quantities.
A parsimonious choice of observables makes it more feasible that $y_{ev}$ can be predicted in all candidate approaches to the nuclear many-body problem.

The program for UQ that we have laid out up to this point in the section could be carried out using ab initio methods and already calibrated chiral-EFT forces and operators. We now discuss a longer-term strategy for refining the prediction of {\NME}. Once it has been established which observables in $y_{ev}$ are strongly correlated with {\NME} the predictive power of the methods $\mathcal{M}_k$ can be improved by  including those members of the dataset $y_{ev}$ in the dataset used to calibrate the nuclear Hamiltonian and decay operators. The parameter vector $\theta$ would then be readjusted within each calculation
$\mathcal{M}_k$. This would open a window for a more refined combined prediction of the different many-body methods. 

One critical challenge to overcome is providing predictions of quantities for various values of $\theta$, $\eta$,  and $\gamma$ for all methods.
Part of this is needed for the integration to compute $p(y|\mathcal{M}_k)$, where MCMC must be leveraged for both the core Hamiltonian parameters $\theta$ as well as the ancillary parameters $\gamma$ and the $0 \nu \beta \beta$ parameter $\eta$.
This problem is amplified in the case of high-dimensional parameter space. 

There are a few statistical and computational tools that can be deployed to resolve this problem.
Firstly, reducing the space of parameters through screening will be critical for each method, thereby including only the parameters that are critical for predicting $y_{ev}$ and {\NME}.
Secondly, emulators (or surrogates) can play a vital role to conduct Bayesian inference from only a few full ab initio evaluations of the matrix element.
Emulators can take on a variety of forms but the function is shared: to provide a computationally inexpensive approximation of  $y_{ev}$ and {\NME} for any value of $\theta$, $\eta$,  and $\gamma$.
 Gaussian process-type emulators exploit smoothness in the computation's response to the parameters.
Other reduced-basis emulators instead build modified, cheaper alternatives to the full ab initio result when it is subject to specific structure~\cite{benner2015survey,hesthaven2015certified,Quarteroni:218966,Benner_2017aa,Benner20201,Benner2020Volume3Applications}.  Variational machine learning methods~\cite{2013arXiv1312.6114K, Bohm:2019hpu, moriconi_high-dimensional_2020} form another path that produce efficient emulators that intrinsically learn the optimal latent parameter space needed for robust interpolation and prediction.  These methods have an advantage over many forms of emulation as they can learn highly non-linear manifolds while still generating a notion of the emulators' internal uncertainty.

One common theme across all methods is they rely on using specific computations at designated parameter combinations to build predictions on other ones.
The amount of computations needed to build an adequate emulator varies across methodologies.
Classical literature on Gaussian process-type emulators suggest computing at ten times the dimension of the parameters, but that suggestion has recently been reconsidered and higher-dimensional parameter spaces perhaps need more computation.
Good emulators will naturally come with their own uncertainty quantification, which is critical for producing valid approximations of $p(y_{ev}|\mathcal{M}_k)$ as well as predictions of $p(M_{0 \nu} | \mathcal{M}_k)$.

As mentioned above,   lower resolution approaches, such as DFT,  can be used to construct emulators of higher resolution ab initio methods. This construction can follow Eq.~(\ref{eq:KOH}) in which $y_{\rm exp}(x;{\theta})$ is  replaced by 
ab initio predictions and  the model discrepancy  $\delta y_{\rm th}$ would model the difference between predictions of ab initio and lower resolution models.

\section{Summary}
\label{sec:summary}
\noindent
Accurate calculation of the nuclear matrix
elements governing neutrinoless double-beta decay, with quantified uncertainty,
is essential for the success of the impressive experimental and theoretical 
worldwide effort in this area \cite{Alarcon:2022ero,Snowmass2}. The purpose of this 
report is  to lay out the challenges to the nuclear-theory community---in regard to both nuclear-physics calculations and uncertainty quantification therein---and map out a path 
for near- and long-term progress. 

It will require a concerted effort in both LQCD and EFT, as well as the 
coupling of these theories, to fully quantify the theoretical 
uncertainties related to the  ${0 \nu \beta \beta}$-decay operator and associated matrix elements at the hadronic level. 
Systematic improvements in nuclear many-body methods  are underway, and should 
be ready to produce a new generation of {\NME} matrix elements in the next 
few years. The complexity of the problem makes the use of multiple 
many-body methods well worthwhile.  Their complementary strengths and 
deficiencies can not only be exploited for validation but also combined through 
Bayesian methods to yield better overall predictions.   

The implementation and application of both QCD/chiral EFT and nuclear many-body  
approaches will require exascale computing resources and beyond.
It will also require an investment in personnel to advance EFT calculations of $0 \nu \beta \beta$ operators and  to develop and maintain the codes 
that implement new theoretical methods. Without such an investment computing time cannot be used
efficiently.  One focus of future work in this area will be ensuring optimal use of the heterogeneous 
architectures that characterize leadership-class computers.  

In addition, the cohesion within and among research groups needs to be strengthened, e.g., through the 
establishment
of joint project resources and inter-institutional collaborations in both pure and computational theory. The first wave of ab initio nuclear matrix 
elements came from the combined efforts of multiple groups, each with its own 
computing resources, and were often obtained at the expense of other groups.
Wait times were long, not only because of a lack of computing resources but also
because certain implementation steps required the work of a single specific (and
busy) person, or because large amounts of data needed to be moved between 
computing systems.  The community needs structures that reduce the severity of these kinds of  bottlenecks.

A crucial feature of this report is the emphasis it places on 
UQ.  Without principled UQ, the usefulness of 
predicted {\NME} values for guiding experimental efforts, interpreting 
measurements, and  assessing new physics will be limited.
At present, few physicists working on the problem of {\NME} make informed choices about UQ, understand the modern UQ glossary, or consider UQ to be an essential part of ``the answer''. 
This situation can be improved 
through coherent inter-disciplinary collaboration
of nuclear physicists with applied mathematicians, statisticians, and computer-science experts. 

Such a collaboration could carry out the concrete, multi-staged, and interwoven program of nuclear-physics and UQ methodological improvements and computations laid out in this report. In concert with continued strong support for the efforts of PIs and research groups working on $0 \nu \beta \beta$ decay, this will make 
the ultimate goal of accurate and precise {\NME} predictions achievable.  


\PRLsep

\noindent
{\it Acknowledgements} We thank Julieta Gruszko for important guidance on the experimental situation in the study of $0 \nu \beta \beta$ decay and for useful comments on this document.
This material is based upon work supported by the National Science Foundation under award number 2004601 of the CSSI program (BAND collaboration), as well as awards 1953111 (Northwestern) and 1913069 (Ohio State), the Alfred P. Sloan foundation and the U.S.\ Department of Energy, Office of Science, Offices of Nuclear Physics and Advanced Scientific Computing Research under award numbers DE-SC0020271 (University of Maryland), together with awards DE-SC0013365, DE-SC0017887, DE-SC0018083  (Michigan State University), DE-SC0004286 (Ohio State University), DE-FG02-03ER41272 (San Diego State University),  DE-FG02-93ER40756 (Ohio University), DE-AC02-06CH11357 (Argonne National Laboratory), DE-FG02-96ER40963 and DE-SC0018223 (University of Tennessee), DE-AC05-00OR22725 (Oak Ridge National Laboratory), DE-SC0022538 (Central Michigan University), DE-AC02-05CH11231 (Lawrence Berkeley National Laboratory), DE-FG02-97ER41019a (University of North Carolina), and  DE-FG02-00ER41132 (Institute for Nuclear Theory), and DE-SC0011090 and DE-SC0021006 (MIT). We especially thank the NSF for funding this Project Scoping Workshop under award number PHY-2226819.

\bibliography{references}
\end{document}